# DIFFnet: Diffusion parameter mapping network generalized for input diffusion gradient schemes and b-values

Juhung Park[1], Woojin Jung[1], Eun-Jung Choi[1], Se-Hong Oh[2], Dongmyung Shin[1], Hongjun An[1] and Jongho Lee[1]

**Author affiliations:**

[1]Laboratory for Imaging Science and Technology, Department of Electrical and Computer Engineering, Seoul National University, Seoul, Korea

[2]Division of Biomedical Engineering, Hankuk University of Foreign Studies, Gyeonggi-do, Korea

**Corresponding Author:**

Jongho Lee, Ph.D

Department of Electrical and Computer Engineering, Seoul National University

Building 301, Room 1008, 1 Gwanak-ro, Gwanak-gu, Seoul, Korea

Tel: 82-2-880-7310

E-mail: jonghoyi@snu.ac.kr

**Abstract**

In MRI, deep neural networks have been proposed to reconstruct diffusion model parameters. However, the inputs of the networks were designed for a specific diffusion gradient scheme (i.e., diffusion gradient directions and numbers) and a specific b-value that are the same as the training data. In this study, a new deep neural network, referred to as DIFFnet, is developed to function as a generalized reconstruction tool of the diffusion-weighted signals for various gradient schemes and b-values. For generalization, diffusion signals are normalized in a q-space and then projected and quantized, producing a matrix (Qmatrix) as an input for the network. To demonstrate the validity of this approach, DIFFnet is evaluated for diffusion tensor imaging ($\text{DIFFnet}_{\text{DTI}}$) and for neurite orientation dispersion and density imaging ($\text{DIFFnet}_{\text{NODDI}}$). In each model, two datasets with different gradient schemes and b-values are tested. The results demonstrate accurate reconstruction of the diffusion parameters at substantially reduced processing time (approximately 8.7 times and 2240 times faster processing time than conventional methods in DTI and NODDI, respectively; less than 4% mean normalized root-mean-square errors (NRMSE) in DTI and less than 8% in NODDI). The generalization capability of the networks was further validated using reduced numbers of diffusion signals from the datasets. Different from previously proposed deep neural networks, DIFFnet does not require any specific gradient scheme and b-value for its input. As a result, it can be adopted as an online reconstruction tool for various complex diffusion imaging.

## INTRODUCTION

Diffusion magnetic resonance imaging (dMRI) non-invasively measures the diffusion characteristics of water molecules and has been widely applied in neuroscience and clinic [1], [2]. In dMRI, various microstructure diffusion models have been developed to extract complex diffusion characteristics [3]-[5]. Among them, diffusion tensor imaging (DTI) [4] and neurite orientation dispersion and density imaging (NODDI) [3] are popular models, measuring water diffusivity and tissue microstructural properties.

When reconstructing a complex microstructure diffusion model (e.g., NODDI), non-linear fitting is often required, costing substantial processing time. For example, the processing time of the NODDI model for a whole-brain dataset takes over 10 hours, limiting real-time processing on the scanner. To amend this limitation, various methods have been proposed to reduce the amount of computation cost [6], [7]. However, the processing time still takes several minutes in spite of decreased accuracy. Hence, further efforts are necessary to reduce the processing time while maintaining accuracy.

Recently, deep learning has been widely applied for the reconstruction and processing of MRI data [8]. A deep neural network trained with a sufficient amount of dataset has been shown to generate highly accurate results at a significantly reduced computational cost [9], [10]. However, when a test dataset has different characteristics (e.g., resolution, signal to noise ratio) from the training dataset, the performance of the deep neural network degrades substantially [11], [12]. This issue of generalization in deep neural networks has been demonstrated to be critical when applying networks for routine practice.

In dMRI reconstruction, neural networks have successfully generated accurate results at a reduced processing time [13]-[15]. In our knowledge, however, these networks proposed so far require a specific diffusion gradient scheme (i.e., specific gradient directions and number of gradients) with specific b-values that are the same as the training data as the input of the network. Such networks require a new training, when input data have a different gradient scheme or b-value, costing a long training time and efforts.

In this study, we present a new deep neural network for diffusion data reconstruction. This deep neural network, referred to as DIFFnet, is generalized for input gradient schemes and b-values by introducing an input matrix. Two DIFFnets are designed and evaluated: $\text{DIFFnet}_{\text{DTI}}$ for DTI and $\text{DIFFnet}_{\text{NODDI}}$ for NODDI. The results of DIFFnet are compared with respect to the conventional fitting methods and the previously proposed neural networks. The source code of DIFFnet is available at https://github.com/SNU-LIST/DIFFnet.

## METHODS

*Deep neural networks: DIFFnet*

The outline of DIFFnet is presented in Fig. 1. For DTI, $\text{DIFFnet}_{\text{DTI}}$ was designed to reconstruct the four DTI model parameters, fractional anisotropy (FA), mean diffusivity (MD), axial diffusivity (AD), and radial diffusivity (RD) from DTI data, which had $n_0$ number of the diffusion gradient directions with a single b-value of $b_0$ s/mm$^2$ (Fig. 1a). For NODDI, $\text{DIFFnet}_{\text{NODDI}}$ generates the three NODDI model parameters, intracellular volume fraction (ICVF), isotropic volume fraction (ISOVF), and orientation dispersion index (ODI) from three-shell diffusion data with three b-values ($b_1$, $b_2$, and $b_3$ s/mm$^2$) and corresponding three numbers ($n_1$, $n_2$, and $n_3$) of diffusion gradient directions (Fig. 1b). The networks were targeted to produce the parameter maps from any reasonable diffusion gradient schemes and b-values.

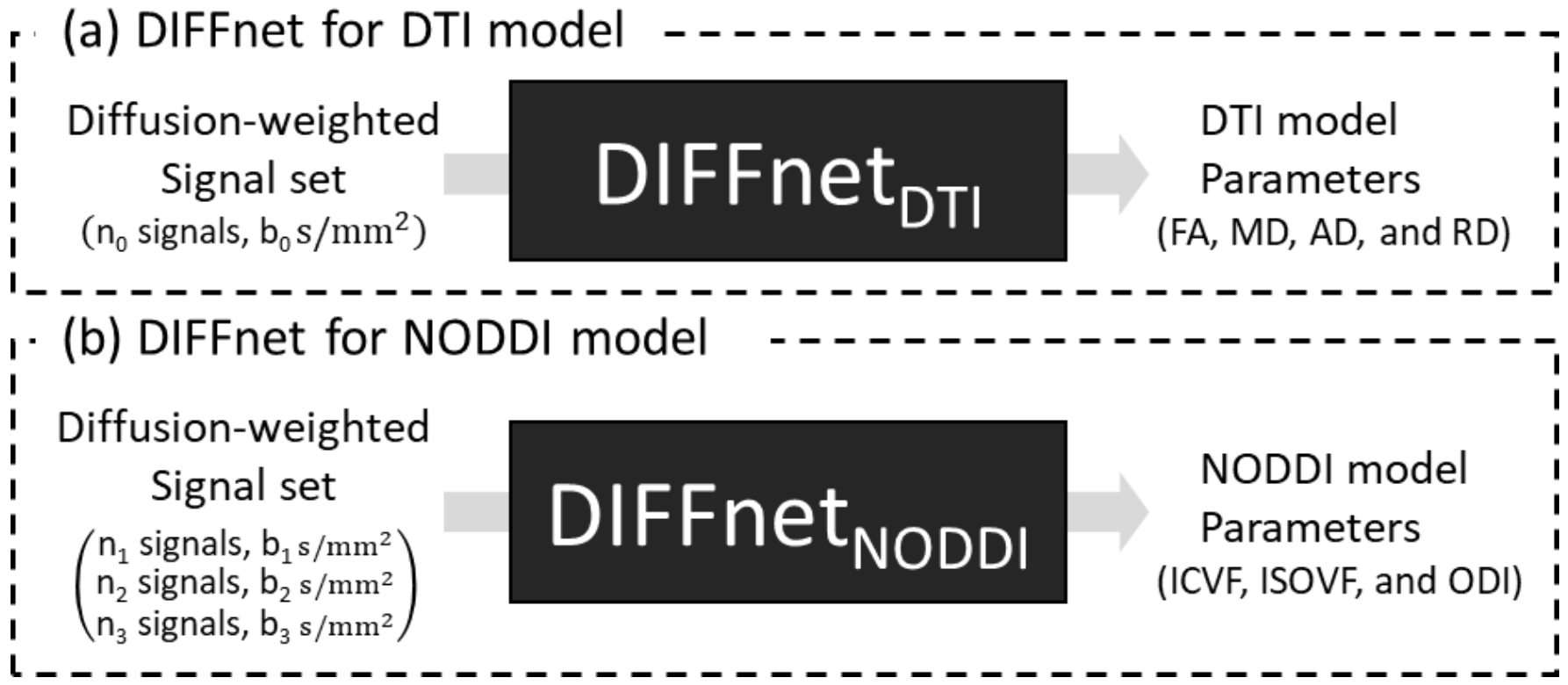

**Figure 1.** Overview of DIFFnet. Two DIFFnets, $\text{DIFFnet}_{\text{DTI}}$ for DTI and $\text{DIFFnet}_{\text{NODDI}}$ for NODDI were designed. (a) $\text{DIFFnet}_{\text{DTI}}$ reconstructed the four DTI parameters, FA, MD, AD, and RD, from DTI data, which had $n_0$ number of diffusion gradient directions with a single b-value of $b_0$ s/mm$^2$. (b) $\text{DIFFnet}_{\text{NODDI}}$ generated the three NODDI parameters, ICVF, ISOVF, and ODI, from three-shell diffusion data with three b-values ($b_1$, $b_2$, and $b_3$ s/mm$^2$) and corresponding three numbers ($n_1$, $n_2$, and $n_3$) of diffusion gradient directions. The networks were targeted to produce the parameter maps from any reasonable diffusion gradient schemes and b-values.

The network structure of DIFFnet was a modified version of the residual neural network, shown in Supplementary Information Fig. S1 [16]. The network was end-to-end trained using one set of diffusion-weighted signals, normalized by the signals with no diffusion weighting, as an input and diffusion model parameters (e.g., FA, MD, AD, and RD or ICVF, ISOVF, and ODI) as a label.

To achieve the generalization for gradient directions and b-values, we introduced an input matrix "Qmatrix", which was a projected and quantized input matrix for the diffusion data. In the first step, a diffusion-weighted signal set was placed in a q-space [17] with q-vectors normalized by the b-value of 1300 s/mm$^2$ for DTI or 2300 s/mm$^2$ for NODDI [18] (Fig. 2a). From this q-space, two different input matrix formats, Qmatrix$_{3D}$ and Qmatrix$_{2D}$, were designed and tested. In Qmatrix$_{3D}$, the q-space was quantized by $q_n$ (Fig. 2c), which was evaluted for five different values (5, 10, 15, 20, and 25), along the three axes, producing a $q_n \times q_n \times q_n$ matrix as the input of the network (Fig. 2b). This matrix was used for both DTI and NODDI. In Qmatrix$_{2D}$ for DTI, the q-space was projected onto xy-, yz-, and xz-planes, then quantized by $q_n \times q_n$, which was also tested for the five different values, producing a $q_n \times q_n \times 3$ matrix (Fig. 2e). For NODDI, the 2D projection was performed for each of the three shells, generating a $q_n \times q_n \times 9$ matrix (Fig. 2f) (see Discussion for using the $q_n \times q_n \times 3$ matrix in NODDI). Consequently, the gradient direction and b-value of each signal were represented by the position of the signal in Qmatrix, allowing various gradient schemes and b-values to be used for the input of DIFFnet.

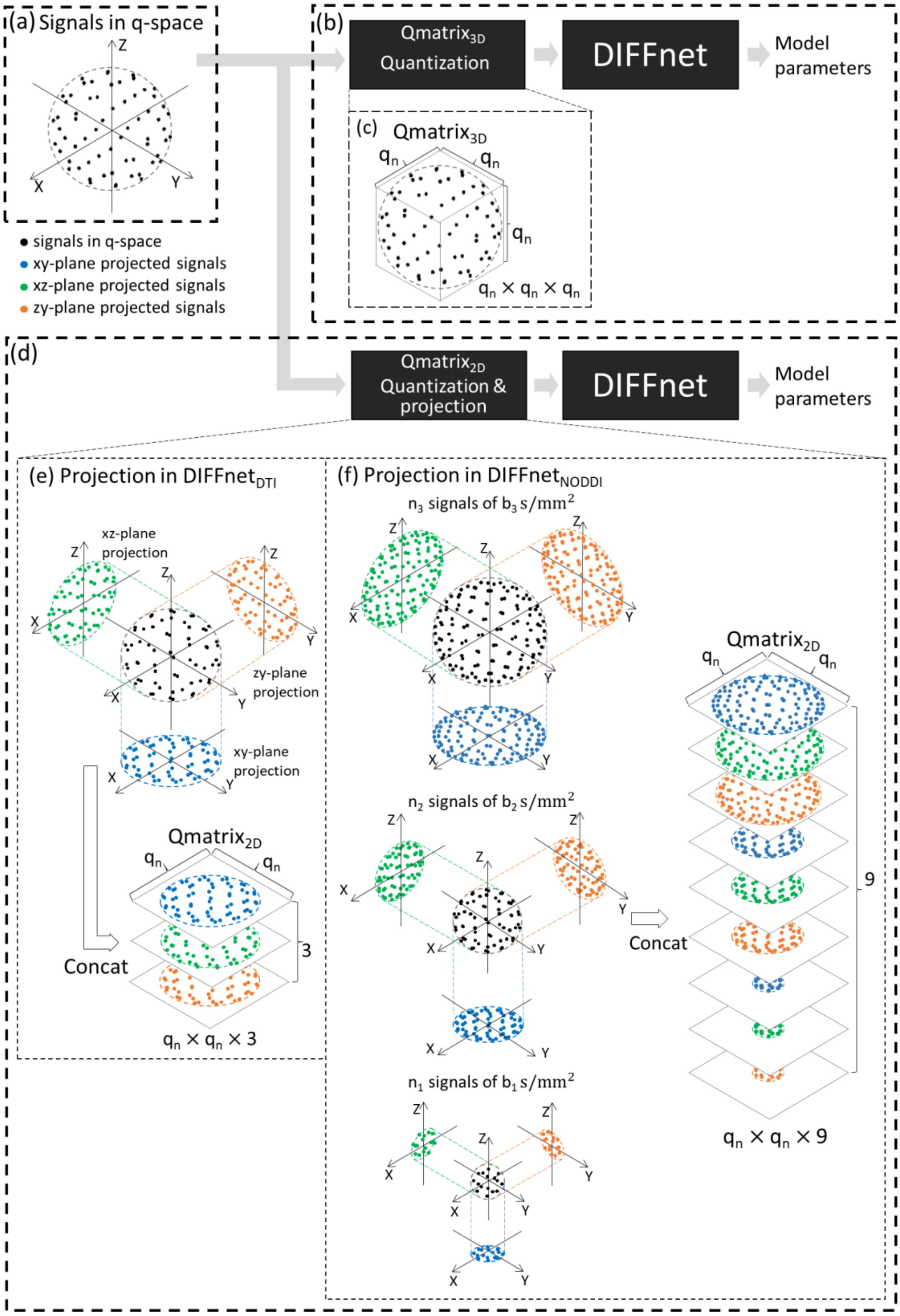

**Figure 2.** Design of Qmatrix for the generalization of diffusion parameter mapping for input diffusion gradient schemes and b-values. (a) A signal set was placed in the q-space with q-vectors normalized by the b-value of 1300 s/mm$^2$ for DTI or 2300 s/mm$^2$ for NODDI. From this q-space, two different input matrix formats, Qmatrix$_{3D}$ and Qmatrix$_{2D}$, were designed. (b and c) For Qmatrix$_{3D}$, the q-space was quantized by $q_n$ along the three axes, producing a $q_n \times q_n \times q_n$ matrix as the input matrix. In $Qmatrix_{2D}$, the design was different for DTI and NODDI. (e) For $Qmatrix_{2D}$ in DTI, the signal set was projected onto the xy- (blue dots), yz- (orange dots), and xz- (green dots) planes. Then, each projected data were quantized by $q_n \times q_n$ and concatenated, producing a $q_n \times q_n \times 3$ matrix. (f) For NODDI, the projection was performed on each shell, generating nine sets of projected planes with a $q_n \times q_n \times 9$ matrix.

*Training dataset generation & training*

The training dataset of DIFFnet was generated solely from Monte-Carlo diffusion simulation. The diffusion characteristic of protons was modeled as a tensor, which had three diffusion coefficients ($d_1$, $d_2$, and $d_3$) with the corresponding eigenvectors ($\vec{e}_1$, $\vec{e}_2$, and $\vec{e}_3$), and was determined by either DTI model or NODDI model. The diffusion tensor of the DTI simulation was determined as follows: First, $d_1$ was randomly chosen between 0 and $3.5\times10^{-3}$ $mm^2s^{-1}$. Subsequently, $d_2$ and $d_3$ were also decided between 0 and $d_1$. The diffusion vector $\vec{e}_1$ was determined at random. The b-value ($b_0$) and number of directions ($n_0$) were chosen in the range of 600 to 1300 $s/mm^2$ and 30 to 80, respectively. In the NODDI simulation, intracellular, extracellular, and cerebrospinal fluid (CSF) compartments were simulated [3]. Protons were allocated to the three compartments following ICVF and ISOVF, which were determined randomly between 0 and 1. Then, the diffusion characteristics were set as follows: In the intracellular compartment, $d_1$ was fixed as the parallel diffusion coefficient $d_{\parallel}$ (= $1.7\times10^{-3}$ $mm^2s^{-1}$), while 0 for the others [19]. For $\vec{e}_1$, Watson distribution was utilized, randomly selecting the mean orientation $\vec{\mu}$ and ODI (0 to 1) [3], [20]. In the extracellular compartment, $d_1$ was set to be the apparent parallel diffusion coefficient $d_{\parallel}'$ whereas the others were set to be the apparent perpendicular diffusion coefficient $d_{\perp}'$, of which $d_{\parallel}'$ and $d_{\perp}'$ are the functions of ODI [3], [21], [22]. It was assumed that $\vec{e}_1$ is the same as $\vec{\mu}$. Lastly, in the CSF compartment, an isotropic diffusion coefficient $d_{iso}$ (= $3.0\times10^{-3}$ $mm^2s^{-1}$) was used for all three diffusion coefficients [3]. The ranges for $b_1$, $b_2$, and $b_3$ were 200 to 400, 500 to 900, 1700 to 2300 $s/mm^2$, respectively. Those for $n_1$, $n_2$, and $n_3$ were 5 to 10, 25 to 50, and 50 to 100, respectively.

A total of $10^6$ protons were generated to create a diffusion-weighted signal. Each proton was assumed to have a unit magnetization, and performed random walks with Gaussian distribution for a time step of Δt (= 0.2 ms). The diffusion simulation was conducted based on a pulsed gradient spin-echo diffusion sequence [2] with TE of 72 ms in DTI and 95 ms in NODDI. For each time step, the phase of each spin magnetization, which was affected by the diffusion gradient, was accumulated. When the simulation reached TE, the average signal of the protons was calculated by the complex sum of all the magnetizations. The signal to noise ratio (SNR) was selected between 30 to 100 by adding Gaussian noise to real and imaginary axes of the signals. The diffusion simulations were performed using MATLAB (MATLAB 2019a, MathWorks Inc., Natick, MA, USA).

For the labels of $\text{DIFFnet}_{\text{DTI}}$, the simulated data were processed using the conventional method [4], [23], generating FA, MD, AD, and RD. In $\text{DIFFnet}_{\text{NODDI}}$, ICVF, ISOVF, and ODI from the conventional method were used as the labels [3]. In both simulations, A total of $10^6$ input signal sets and label model parameter pairs were produced for the training.

The training was performed on a GPU workstation (NVIDIA GeForce GTX 1080Ti GPU [NVIDIA Corp., Santa Clara, CA] with Intel Xeon CPU E5-2630 v3 at 2.40GHz [Intel Corp., Santa Cruz, CA]) using TensorFlow [24]. The initial weights of the convolutional kernel were set by Xavier initializer [25]. The batch size was 100. The loss function was defined as mean-squared-error and Adam optimizer was utilized [26]. The initial learning rate was $10^{-3}$ with a decaying factor of 0.87 for each epoch. The training process was stopped after 50 epochs.

*MRI data acquisition & post-processing*

For the validation of $\text{DIFFnet}_{\text{DTI}}$ and $\text{DIFFnet}_{\text{NODDI}}$, two types of *in-vivo* data ($\text{Dataset}_{\text{DTI-A}}$ and $\text{Dataset}_{\text{DTI-B}}$ for DTI; $\text{Dataset}_{\text{NODDI-A}}$ and $\text{Dataset}_{\text{NODDI-B}}$ for NODDI) were used. $\text{Dataset}_{\text{DTI-A}}$ and $\text{Dataset}_{\text{NODDI-A}}$ were from Jung, et al. [27]. $\text{Dataset}_{\text{DTI-B}}$ and $\text{Dataset}_{\text{NODDI-B}}$ were obtained for this study to test the effects of a different diffusion gradient scheme and had different gradient directions and b-values from $\text{Dataset}_{\text{DTI-A}}$ and $\text{Dataset}_{\text{NODDI-A}}$. All subjects (10 subjects) were scanned with a 3T MRI system (Tim Trio, SIEMENS, Erlangen, Germany) using a 32-channel phased-array head coil. The study was approved by the institutional review board.

For $\text{Dataset}_{\text{DTI-A}}$, single-shell data (b = 700 s/mm$^2$ with 32 directions; b = 0 s/mm$^2$ with 13 averages) were acquired using a single-shot spin-echo echo-planar-imaging (SE-EPI) sequence. The scan parameters were TR/TE = 4000/95 ms, FOV = 192×192 mm$^2$, voxel size = 2×2 mm$^2$, slice thickness = 2 mm, multi-band factor = 2, GRAPPA factor = 2, and partial Fourier = 6/8. The dataset had five subjects.

For $Dataset_{NODDI-A}$, the same data as in $Dataset_{DTI-A}$ were used, with an addition of b = 300 s/mm$^2$ in 8 directions and b = 2000 s/mm$^2$ in 64 directions data.

For $Dataset_{DTI-B}$, five healthy subjects were scanned. Single-shell data (b = 1000 s/mm$^2$ with 30 directions; b = 0 s/mm$^2$ with 4 averages) were acquired using a single-shot SE-EPI sequence. The scan parameters were TR/TE = 3500/72 ms, FOV = 256×256 mm$^2$, voxel size = 2×2 mm$^2$, slice thickness = 2 mm, multi-band factor = 2, GRAPPA factor = 3, and partial Fourier = 6/8.

For $Dataset_{NODDI-B}$, five healthy subjects were scanned. Three-shell data (b = 300 s/mm$^2$ with 8 directions; b = 700 s/mm$^2$ with 30 directions; b = 2000 s/mm$^2$ with 60 directions; b = 0 s/mm$^2$ with 13 averages) were acquired using a single-shot SE-EPI sequence. The scan parameters were TR/TE = 3000/105 ms, FOV = 240×240 mm$^2$, voxel size = 1.5×1.5 mm$^2$, slice thickness = 2 mm, GRAPPA factor = 2, and partial Fourier = 6/8.

To compensate for EPI geometric distortion, a reversed-phase encoding direction scan with b = 0 s/mm$^2$ was acquired for all datasets. The b-values and the number of gradient vectors of all datasets are summarized in Table I (see Supplementary Information Table SI for the directions of the gradient vectors). All datasets had different gradient vector directions.

All datasets were processed as follows: TOPUP and EDDY (FSL, FMRIB, Oxford, UK) [28] were used for geometric distortion. A brain tissue mask was generated from the magnitude image with b = 0 s/mm$^2$ using BET (FSL, Oxford, UK) [29]. For $Dataset_{DTI-A}$ and $Dataset_{DTI-B}$, the DTI parameters were reconstructed by least-square-fitting as references [30], [31]. For $Dataset_{NODDI-A}$ and $Dataset_{NODDI-B}$, the NODDI parameters were reconstructed by conventional NODDI as references [3]. Additionally, the NODDI parameters were reconstructed using accelerated microstructure imaging via convex optimization (AMICO), which is commonly utilized for NODDI reconstruction because of computational efficiency [32], [33].

| | | b-value (s/mm$^2$) | Number of directions |
|---|---|---|---|
| DTI | $Dataset_{DTI\text{-}A}$ | 700 | 32 |
| | $Dataset_{DTI\text{-}B}$ | 1000 | 30 |
| NODDI | $Dataset_{NODDI\text{-}A}$ | 300 | 8 |
| | | 700 | 32 |
| | | 2000 | 64 |
| | $Dataset_{NODDI\text{-}B}$ | 300 | 8 |
| | | 700 | 30 |
| | | 2000 | 60 |

**Table 1.** List of the b-values and the numbers of diffusion directions in test datasets. The directions of the gradient vectors are reported in Supplementary information Table S9.

*Evaluation*

To determine the quantization size of Qmatrix ($q_n$ = 5, 10, 15, 20, and 25 for both $Qmatrix_{3D}$ and $Qmatrix_{2D}$), normalized root-mean-square-errors (NRMSEs) were calculated in the brain mask with respect to the reference maps. The data processing time of each quantization size was also computed.

After deciding the optimum quantization size, the performance of DIFFnet was compared with a previously proposed neural network [14], which utilized multi-layer perceptron (MLP). Two MLPs were designed: $MLP_{DTI}$ having five fully connected layers with 32, 256, 256, 256, and 4 neurons for DTI, and $MLP_{NODDI}$ having five fully connected layers with 104, 400, 400, 400, and 3 neurons for NODDI. For the input, 32 diffusion signals were used whereas it was increased to 104 signals for $MLP_{NODDI}$. If a test dataset had less than 32 or 104 signals, the rest was zero-padded. For the training of MLPs, datasets were generated using our Monte-Carlo diffusion simulation with the gradient scheme the same as $Dataset_{DTI\text{-}A}$ or $Dataset_{NODDI\text{-}A}$. All the other training parameters and procedures were the same as those in DIFFnet.

All the processing methods (conventional fitting, DIFFnet, and MLP, additionally AMICO in NODDI) were evaluated by the four datasets. The data processing time for each method was measured. NRMSEs were estimated in the brain mask with respect to the reference maps. A Wilcoxon rank-sum test was performed for NRMSEs between DIFFnet and MLP or between DIFFnet and AMICO for each of the model parameters. For statistical significance, a p-value threshold was set to be 0.05.

To further demonstrate the generalization capability of DIFFnet for a various number of gradient directions, DIFFnet was evaluated using five different numbers of the gradient directions in all four datasets (see Supplementary Information S1 for the numbers of gradient directions and the criteria of selecting gradient directions). The performance of DIFFnet was compared with the conventional methods. Additionally, $DIFFnet_{NODDI}$ was evaluated by a two-shell NODDI protocol, which is also commonly used in practice, using six different numbers of the gradient directions of $Dataset_{NODDI\text{-}A}$ and $Dataset_{NODDI\text{-}B}$.

## RESULTS

When we investigated the effects of the quantization in the q-space using a range of $q_n$, the results reveal the minimum mean NRMSEs at $q_n$ of 20 in $\text{Qmatrix}_{2D}$ and 15 in $\text{Qmatrix}_{3D}$ (Table II). Between the two results, they show similar mean NRMSEs (no statistical difference), but the processing time of $\text{Qmatrix}_{2D}$ is faster than $\text{Qmatrix}_{3D}$ (13.3 times in $\text{DIFFnet}_{DTI}$ and 13.0 times in $\text{DIFFnet}_{NODDI}$). Hence, $\text{Qmatrix}_{2D}$ with $q_n$ of 20 is chosen as the default format for DIFFnet hereafter.

| | $q_n$ | NRMSE (%) | | Processing time (seconds) | |
|---|---|---|---|---|---|
| | | $\text{DIFFnet}_{DTI}$ | $\text{DIFFnet}_{NODDI}$ | $\text{DIFFnet}_{DTI}$ | $\text{DIFFnet}_{NODDI}$ |
| $\text{Qmatrix}_{2D}$ | 5 | 4.35 ± 2.01 | 8.65 ± 4.01 | 11.0 ± 0.6 | 11.1 ± 0.7 |
| | 10 | 2.70 ± 1.65 | 6.18 ± 2.95 | 15.2 ± 0.8 | 15.2 ± 1.1 |
| | 15 | 2.09 ± 1.37 | 5.10 ± 2.34 | 20.2 ± 1.4 | 20.7 ± 1.5 |
| | 20 | 1.88 ± 1.35 | 5.03 ± 2.46 | 26.7 ± 1.6 | 27.8 ± 1.9 |
| | 25 | 2.12 ± 1.43 | 5.22 ± 2.37 | 37.6 ± 2.0 | 39.0 ± 1.9 |
| $\text{Qmatrix}_{3D}$ | 5 | 3.41 ± 2.08 | 7.33 ± 3.90 | 24.8 ± 1.0 | 25.2 ± 1.1 |
| | 10 | 2.38 ± 1.51 | 5.56 ± 2.62 | 120 ± 7 | 121 ± 8 |
| | 15 | 1.85 ± 1.29 | 4.93 ± 2.11 | 356 ± 21 | 362 ± 24 |
| | 20 | 1.93 ± 1.45 | 4.98 ± 2.61 | 774 ± 34 | 785 ± 40 |
| | 25 | 2.25 ± 1.63 | 5.21 ± 2.78 | 1597 ± 77 | 1638 ± 89 |

**Table 2**. Mean NRMSE and processing time measured using a range of $q_n$, which is the number of quantization, in $\text{Qmatrix}_{2D}$ and $\text{Qmatrix}_{3D}$. $\text{Qmatrix}_{2D}$ with $q_n$ of 20 is chosen as the default format for DIFFnet.

In Fig. 3, the DTI maps of the two datasets with different gradient schemes are reconstructed by the conventional fitting, $\text{DIFFnet}_{DTI}$, and MLP trained with the gradient scheme of $\text{Dataset}_{DTI\text{-}A}$. $\text{DIFFnet}_{DTI}$ generates highly accurate parameter maps with respect to those using the conventional fitting in both datasets (NRMSEs of FA: 3.73 ± 0.52%, MD: 0.52 ± 0.07%, AD: 1.79 ± 0.25%, and RD: 0.96 ± 0.08% in $\text{Dataset}_{DTI\text{-}A}$, FA: 3.89 ± 0.31%, MD: 0.67 ± 0.07%, AD: 1.97 ± 0.18%, and RD: 1.13 ± 0.07% in $\text{Dataset}_{DTI\text{-}B}$). The error maps also confirm little difference between the two maps (Fig. 3). The mean processing time of DIFFnet is measured to be faster than that of the conventional fitting (26.7 ± 1.6 s in $\text{DIFFnet}_{DTI}$; 46.1 ± 3.1 s in conventional fitting). When MLP reconstructs $\text{Dataset}_{DTI\text{-}A}$, which has the same gradient scheme as in the MLP training, MLP generates highly accurate parameter maps, showing similar NRMSEs to DIFFnet (NRMSEs of FA: 3.45 ± 0.59%, MD: 0.54 ± 0.06%, AD: 1.76 ± 0.29%, and RD: 0.94 ± 0.08%; no statistical difference). On the other hand, MLP fails to reconstruct $\text{Dataset}_{DTI\text{-}B}$, which has a different gradient scheme, reporting significantly larger NRMSEs than DIFFnet (NRMSEs of FA: 27.52 $\pm$ 9.21%, MD: 12.93 $\pm$ 4.38%, AD: 18.96 $\pm$ 3.56%, and RD: 16.11 $\pm$ 5.04%; Wilcoxon rank-sum test results: $p$ = 0.008 for FA, p = 0.008 for MD, p = 0.008 for AD and p = 0.008 for RD). The mean processing time of MLP is 10.3 $\pm$ 0.57 s.

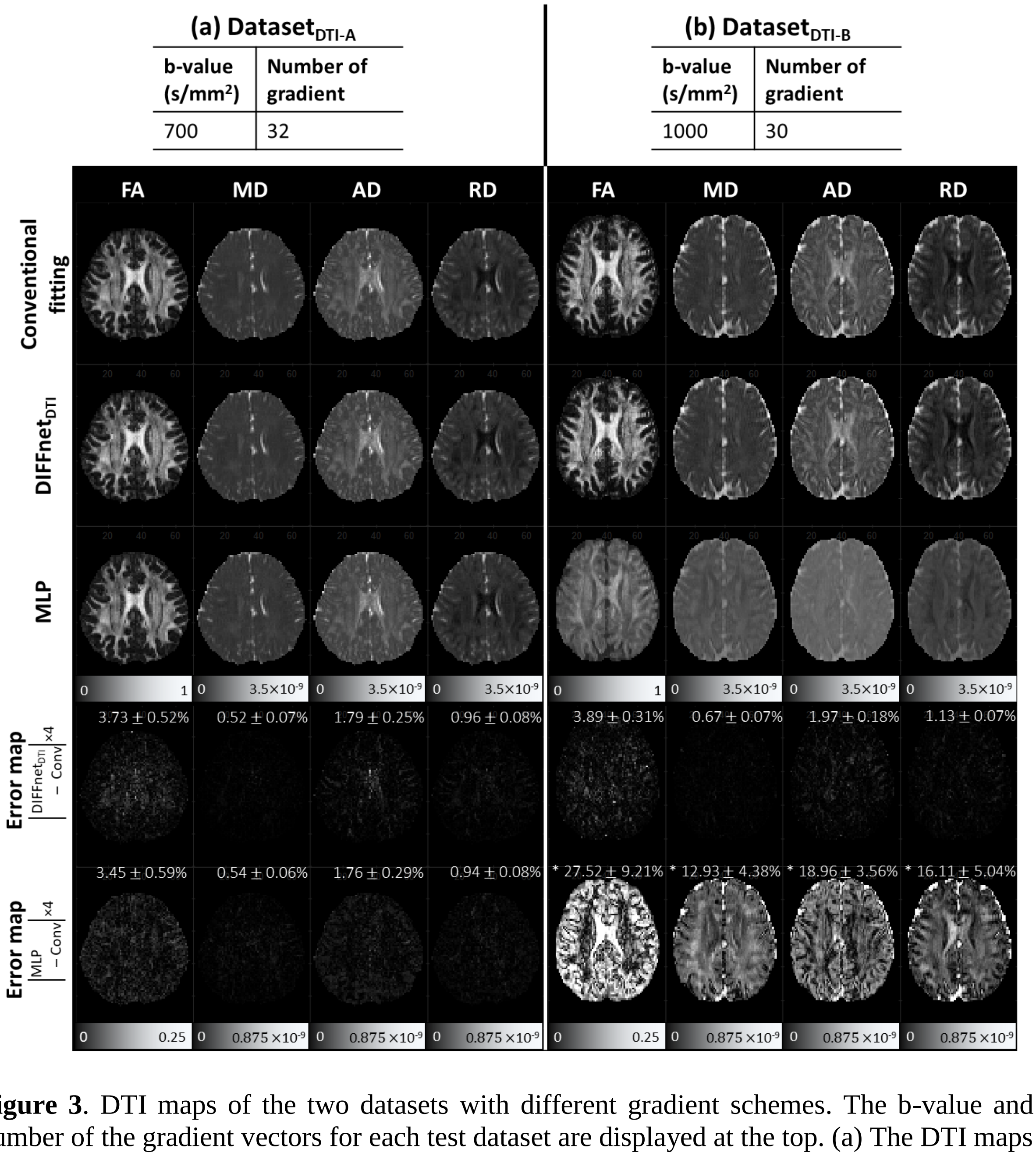

**Figure 3**. DTI maps of the two datasets with different gradient schemes. The b-value and number of the gradient vectors for each test dataset are displayed at the top. (a) The DTI maps of Dataset$_{DTI-A}$ (first to fourth columns) reconstructed by least-square-fitting (first row), DIFFnet$_{DTI}$ (second row), and MLP (third row) are shown. The error maps of DIFFnet$_{DTI}$ (fourth row) and MLP (last row) are also included (display range is reduced by a factor of 4 for visualization). (b) The DTI maps of Dataset$_{DTI-B}$ (fifth to last columns) reconstructed by least-square-fitting (first row), DIFFnet$_{DTI}$ (second row), and MLP (third row) are displayed along

with the errors maps of $DIFFnet_{DTI}$ (fourth row) and MLP (last row). The NRMSE is shown at the top of each error map (* denotes a statistically significant difference between the NRMSEs of $DIFFnet_{DTI}$ and MLP).

In the reconstruction of the NODDI maps, $DIFFnet_{NODDI}$ successfully generates the parameter maps of the two datasets with different gradient schemes (Fig. 4; NRMSEs of ICVF: 3.95 ± 0.21%, ISOVF: 3.70 ± 0.46%, and ODI: 7.96 ± 0.46% in $Dataset_{NODDI\text{-}A}$, ICVF: 3.59 ± 0.34%, ISOVF: 3.51 ± 0.27%, and ODI: 7.82 ± 0.34% in $Dataset_{NODDI\text{-}B}$). Compared to the results of the AMICO reconstruction (NRMSEs of ICVF: 6.67 ± 0.45%, ISOVF: 7.28 ± 0.91%, and ODI: 8.77 ± 1.07% in $Dataset_{NODDI\text{-}A}$, ICVF: 5.81 ± 0.35%, ISOVF: 7.14 ± 0.51%, and ODI: 8.86 ± 0.93% in $Dataset_{NODDI\text{-}B}$), those of $DIFFnet_{NODDI}$ show lower NRMSEs (Wilcoxon rank-sum test results: $p$ = 0.008 for ICVF, $p$ = 0.008 for ISOVF, and $p$ = 0.095 for ODI in $Dataset_{NODDI\text{-}A}$; $p$ = 0.008 for ICVF, $p$ = 0.008 for ISOVF, and $p$ = 0.150 for ODI in $Dataset_{NODDI\text{-}B}$). This tendency can be confirmed in the error maps, which show higher errors in the results of AMICO than $DIFFnet_{NODDI}$, particularly in ICVF and ISOVF (Fig. 4). Another advantage of $DIFFnet_{NODDI}$ is reconstruction time. Compared to AMICO and NODDI, $DIFFnet_{NODDI}$ reveals approximately 8.7 times and 2240 times faster processing time, respectively (27.8 ± 1.4 s in $DIFFnet_{NODDI}$; 242.5 ± 11.8 s in AMICO; 17.3 ± 0.8 h in NODDI). When the performance of $DIFFnet_{NODDI}$ is compared with MLP, similar trends to those in DTI are observed. For $Dataset_{NODDI\text{-}A}$, which has the same gradient scheme as in the MLP training, MLP results show similar NRMSEs to those of $DIFFnet_{NODDI}$ (ICVF: 3.88 ± 0.31%, ISOVF: 3.68 ± 0.44%, and ODI: 7.55 ± 0.42%; no statistical difference). However, MLP fails to reconstruct $Dataset_{NODDI\text{-}B}$, which has a different gradient scheme, reporting significantly larger NRMSEs (NRMSEs of ICVF: 12.91 ± 3.61%, ISOVF: 10.56 ± 2.25%, and ODI: 42.18 ± 8.10%, Wilcoxon rank-sum test results: $p$ = 0.008 for ICVF, p = 0.008 for ISOVF, and p = 0.008 for ODI). The mean processing time of MLP is 13.1 ± 0.94 s.

**(a) Dataset$_{NODDI\text{-}A}$**

| b-value (s/mm²) | Number of gradient |
|---|---|
| 300 | 8 |
| 700 | 32 |
| 2000 | 64 |

**(b) Dataset$_{NODDI\text{-}B}$**

| b-value (s/mm²) | Number of gradient |
|---|---|
| 300 | 8 |
| 700 | 30 |
| 2000 | 60 |

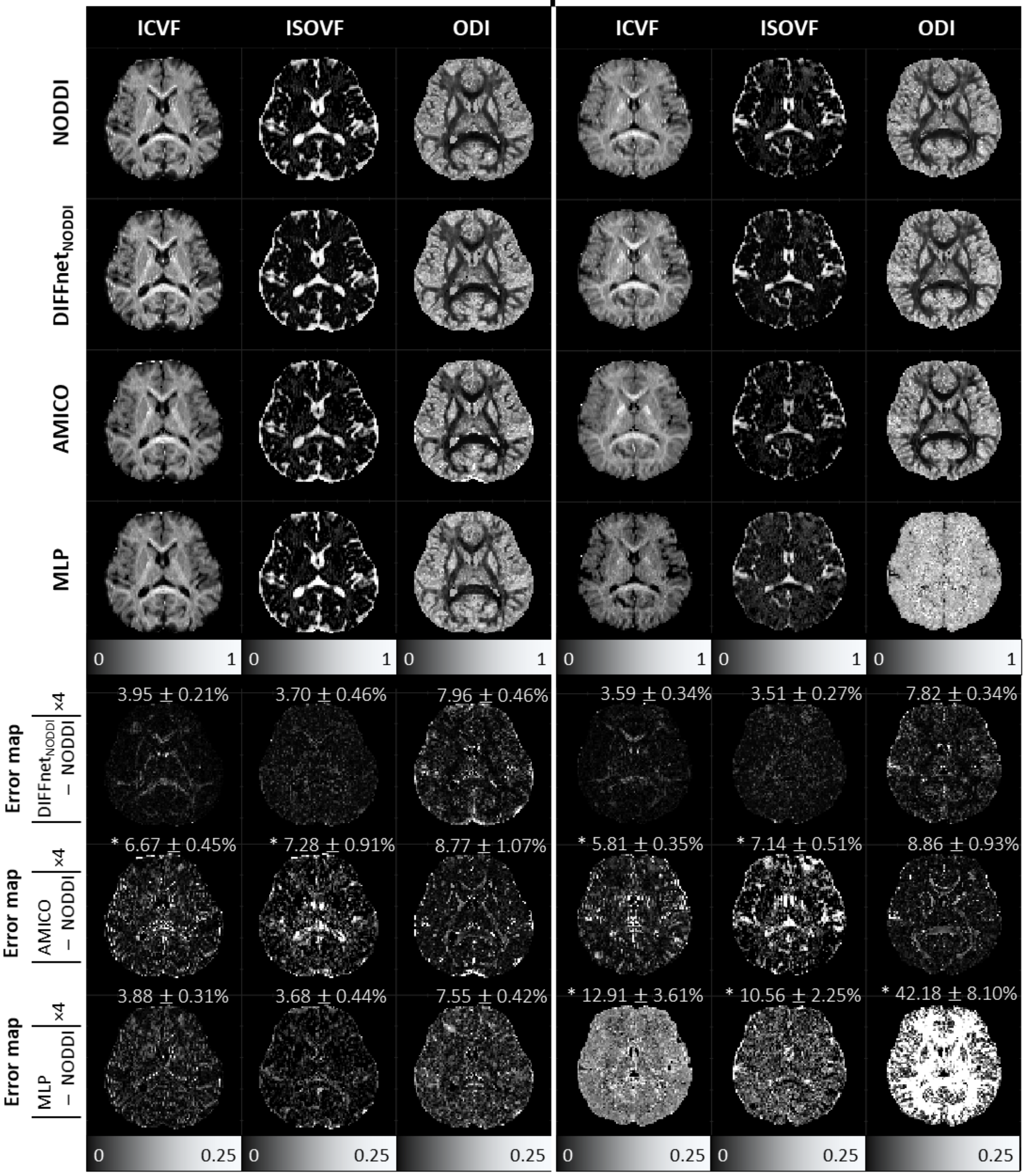

**Figure 4.** NODDI maps of the two datasets with different gradient schemes. The b-values and number of gradient vectors for each test dataset are displayed at the top. (a) The NODDI maps of Dataset$_{NODDI\text{-}A}$ (first to third columns) reconstructed by NODDI (first row), DIFFnet$_{NODDI}$ (second row), AMICO (third row), and MLP (fourth row) are shown. The error maps of DIFFnet$_{NODDI}$ (fifth row), AMICO (sixth row), and MLP (last row) are also included (display range is reduced by a factor of 4). (b) The NODDI maps of Dataset$_{NODDI\text{-}B}$ (fourth to last columns) reconstructed by NODDI (first row), DIFFnet$_{NODDI}$ (second row), AMICO (third

row), and MLP (fourth row) are displayed along with the error maps of $\text{DIFFnet}_{\text{NODDI}}$ (fifth row), AMICO (sixth row), and MLP (last row). The NRMSE is shown at the top of each error map (* denotes there is a statistically significant difference in NRMSEs).

When DIFFnets are further tested using the smaller numbers of the gradient directions for all the four datasets, DIFFnets successfully reconstructed the parameter maps of DTI and NODDI (Supplementary Information Fig. S2, S3, S4, and S5). Additionally, $\text{DIFFnet}_{\text{NODDI}}$ successfully reconstructed NODDI results from the two-shell NODDI dataset (Supplementary Information Fig. S6 and S7), consolidating the generalization capability of DIFFnet.

## DISCUSSION AND CONCLUSION

In this study, we developed a deep neural network, DIFFnet, to reconstruct the diffusion model parameters from diffusion-weighted signals. Unlike previously proposed deep neural networks [13]-[15], DIFFnet was targeted to generate the parameter maps from various gradient schemes and b-values. For the generalization of the input signals, Qmatrix was introduced via q-space projection and quantization. The performance of DIFFnet was evaluated in two diffusion models, DTI and NODDI, with two datasets in each model. The results of DIFFnet demonstrated successful reconstruction, differentiating it from MLP. In the NODDI reconstruction results, DIFFnet outperformed AMICO, reporting higher accuracy. The processing time of DIFFnet was less than 30s, suggesting it can be used for online reconstruction.

In our diffusion simulation, the b-values and number of gradient directions were chosen to include commonly used scan protocols (DTI: b = 600 to 1000 s/mm$^2$, 30 to 32 gradient directions; NODDI: b = 300 to 2000 s/mm$^2$, 90 to 104 gradient directions) [27], [35]-[37]. Similarly, the normalization factors for Qmatrix (b = 1300 s/mm$^2$ in DTI and b = 2300 s/mm$^2$ in NODDI) were large enough to cover the b-values in the commonly used protocols. Nevertheless, for data with a higher b-value, additional simulation data with the corresponding b-value can be added for the training data.

In the Qmatrix of NODDI, the projection was performed for each shell, generating a $q_n \times q_n \times 9$ matrix. When this design was compared with the Qmatrix projected for all shells (i.e., $q_n \times q_n \times 3$ matrix), our results showed higher accuracy (NRMSEs of ICVF: 3.95 $\pm$ 0.21% for $q_n \times q_n \times 9$ vs 4.23 $\pm$ 0.86% for $q_n \times q_n \times 3$; all the other parameters presented similar trends). This inferior performance of the $q_n \times q_n \times 3$ matrix may be explained by the different intensity ranges between the signals with different b-values. Low-intensity signals from the high b-value diffusion signals may be inaccurately processed.

When comparing $q_n$, the largest $q_n$ (= 25) provided the highest resolution in the q-space. However, the results showed the lowest NRMSEs when $q_n$ of 20 in Qmatrix$_{2D}$ and 15 in Qmatrix$_{3D}$. This result may be explained by the size of the convolutional kernel in DIFFnet, which utilized 7 × 7 or 7 × 7 × 7 at the first convolutional layer. This size limit in the kernel may not be sufficient for $q_n$ larger than 20 (or 15).

In our DTI simulation, only $d_1$ was assumed to be the largest among the three diffusion coefficients. This did not lose generality despite the common assumption of $d_1 > d_2 > d_3$.

The computational times for conventional DTI and NODDI including AMICO were calculated using CPU processing whereas that of DIFFnet using GPU processing. Hence the comparison was not fair. The use of GPU for NODDI reconstruction, however, has not been demonstrated.

In DIFFnet, two diffusion models, DTI and NODDI, were chosen as exemplary diffusion models. Since our approach of using Qmatrix is general for any diffusion imaging, it can be applied to other diffusion models.

**Acknowledgments**

This research was supported by the National Research Foundation of Korea (NRF-2018R1A4A1025891 and NRF-2017M3C7A1047864), the Institute of New Media and Communications, and the Institute of Engineering Research at Seoul National University.

## Supplementary Information

Supplementary information 1.

To investigate the effects of the number of gradient directions, DIFFnet and the conventional methods were evaluated using five different numbers of gradient directions in all four datasets. For $DIFFnet_{DTI}$, 12, 16, 20, 24 and 28 gradient directions with b = 700 s/mm$^2$ were tested in $Dataset_{DTI-A}$, and 10, 14, 18, 22 and 26 gradiet directions with b = 1000 s/mm$^2$ were tested in $Dataset_{DTI-B}$. For $DIFFnet_{NODDI}$, 39, 52, 65, 78, and 91 gradient directions (3, 4, 5, 6, and 7 for b = 300 s/mm$^2$; 12, 16, 20, 24, and 28 for b = 700 s/mm$^2$; 24, 32, 40, 48, and 56 for b = 2000 s/mm$^2$) were used in $Dataset_{NODDI-A}$, and 33, 46, 59, 72, and 85 gradient directions (3, 4, 5, 6, and 7 for b = 300 s/mm$^2$; 10, 14, 18, 22, and 26 for b = 700 s/mm$^2$; 20, 28, 36, 44, and 52 for b = 2000 s/mm$^2$) were used in $Dataset_{NODDI-B}$. When selecting gradient directions, a combination that had the lowest condition number was chosen [4]. NRMSEs were estimated based on the reference maps, which utilized the full the gradient directions.

Additionally, $DIFFnet_{NODDI}$ was evaluated using a two-shell protocol. In $Dataset_{NODDI-A}$, 36, 48, 60, 72, 84, and 96 directions (12, 16, 20, 24, 28, and 32 for b = 700 s/mm$^2$; 24, 32, 40, 48, 56, and 64 for b = 2000 s/mm$^2$) were tested. In $Dataset_{NODDI-B}$, 30, 42, 54, 66, 78, and 90 directions (10, 14, 18, 22, 26, and 30 for b = 700 s/mm$^2$; 20, 28, 36, 44, 52, and 60 for b = 2000 s/mm$^2$) were tested.

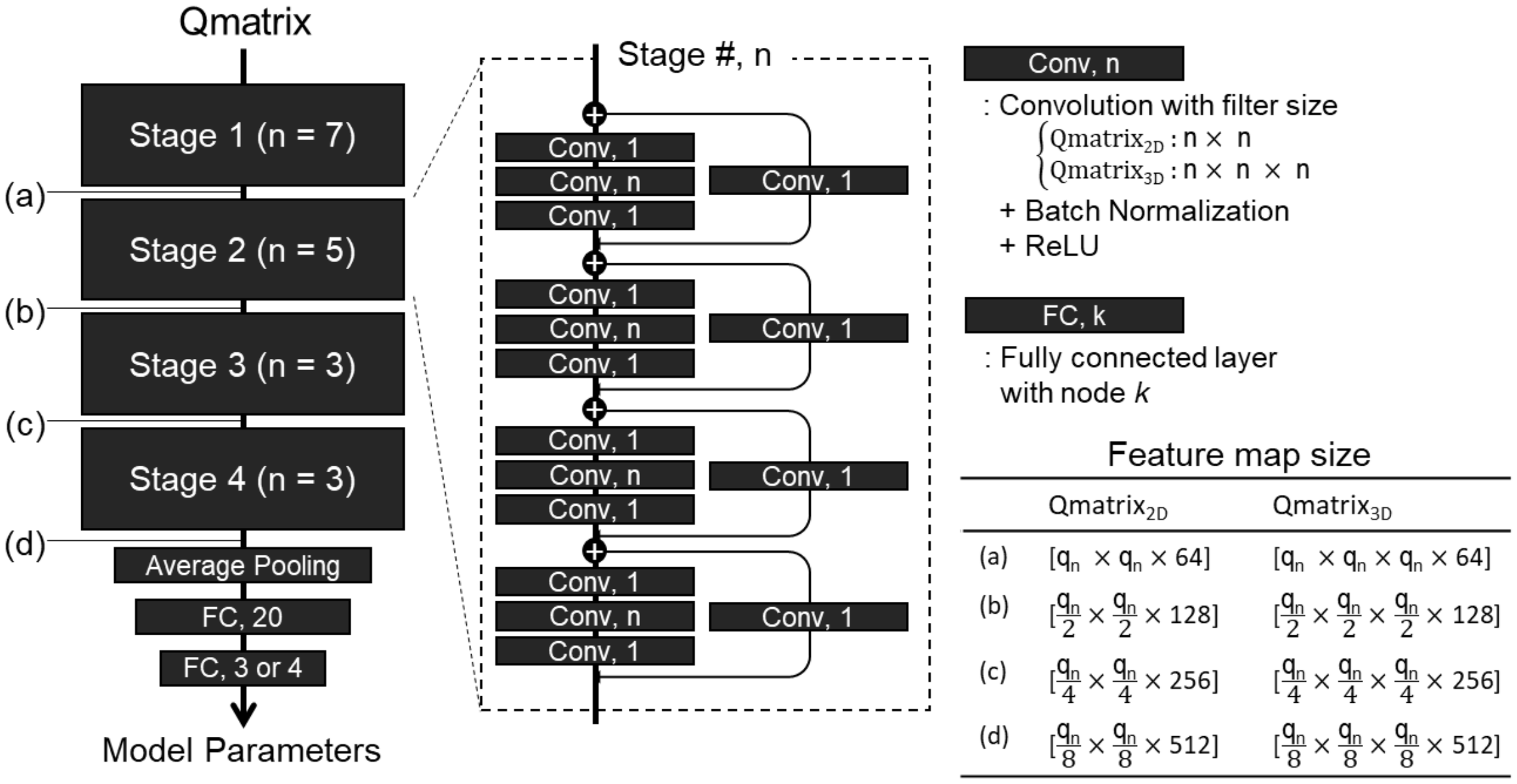

**Fig. S1**. Detailed structure of DIFFnet. A modified version of the residual neural network was utilized. The network consisted of five stages, followed by one averaging pooling layer and two fully connected layers. Each stage contained 13 convolutional layers and 4 skip connections [1]. Each convolutional layer was followed by batch normalization [2] and leaky ReLU (alpha = 0.1) [3]. The first fully connected layer had 20 nodes, and the second fully connected layer had three and four nodes in DTI and NODDI, respectively.

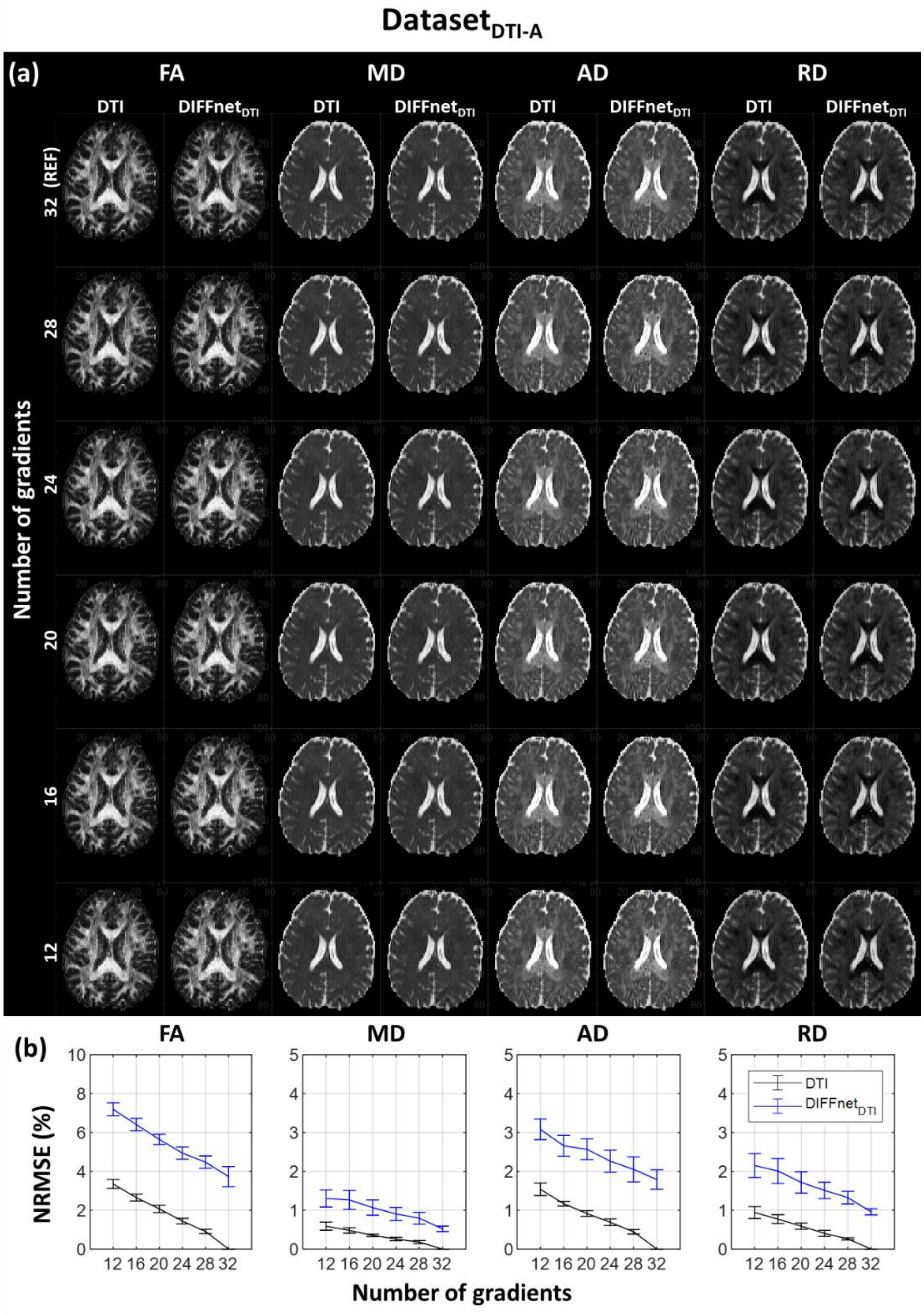

**Fig. S2**. (a) DTI parameter maps of Dataset$_{DTI-A}$ reconstructed by the least-square-fitting and DIFFnet$_{DTI}$ using the five different numbers of the diffusion gradients. (b) NRMSEs in the DIFFnet$_{DTI}$ and least-square-fitting results. The reference for NRMSE was the least-square-fitting results using the 32 diffusion gradient directions.

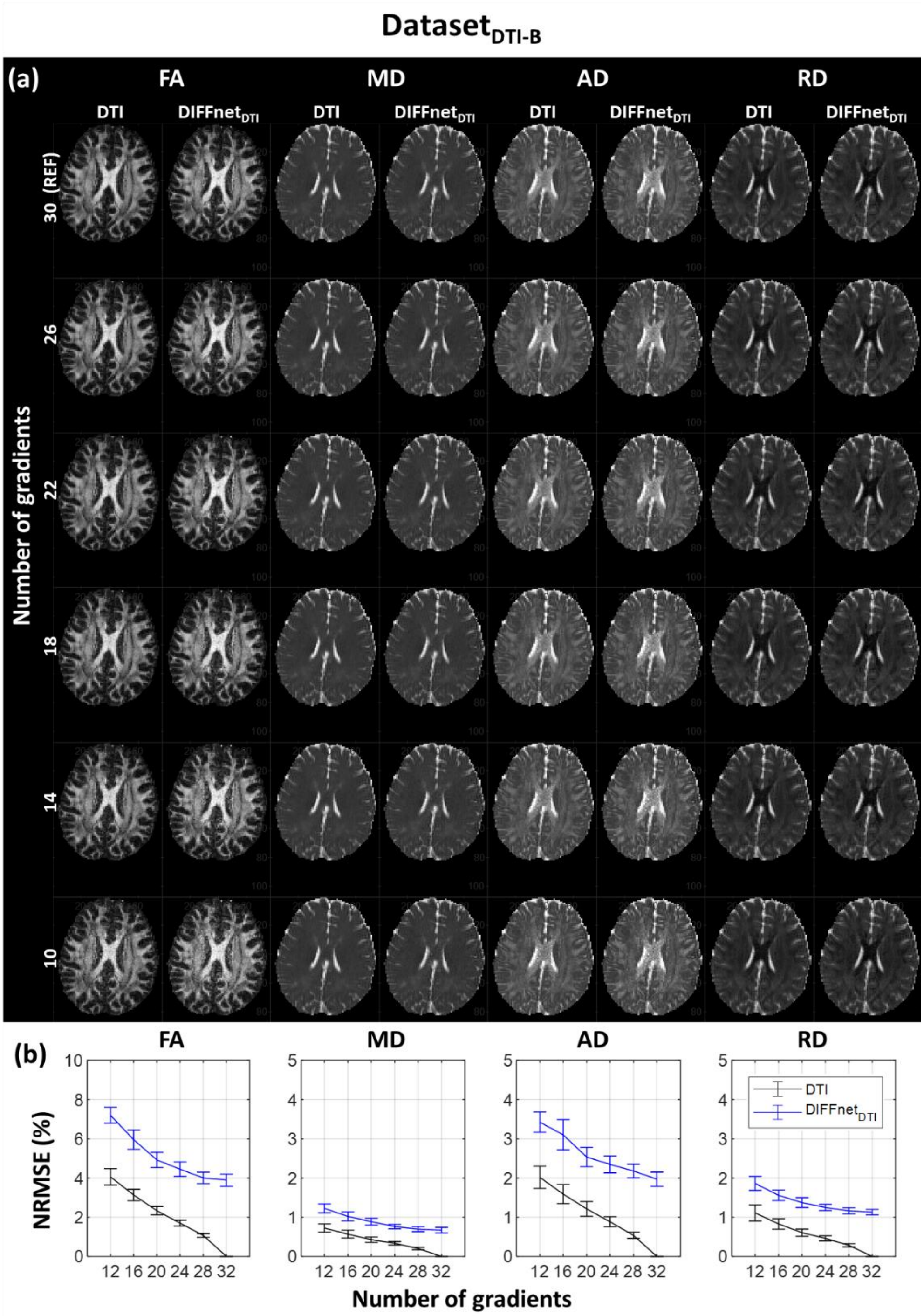

**Fig S3**. (a) DTI parameter maps of Dataset$_{DTI-B}$ reconstructed by the least-square-fitting and DIFFnet$_{DTI}$ using the five different numbers of the diffusion gradients. (b) NRMSEs in the DIFFnet$_{DTI}$ and least-square-fitting results. The reference for NRMSE was the least-square-fitting results using the 30 diffusion gradient directions.

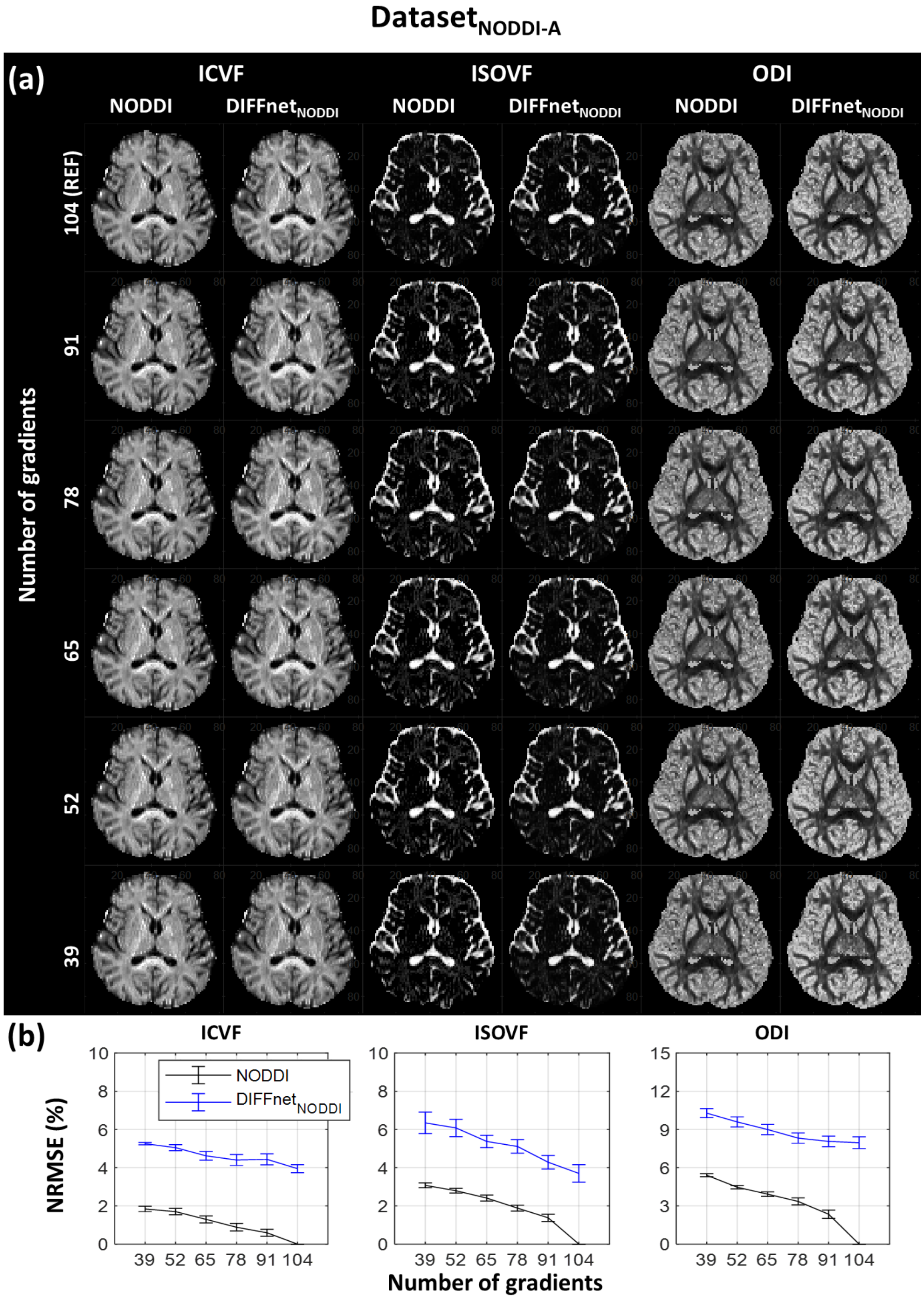

**Fig S4**. (a) NODDI parameter maps of Dataset$_{NODDI\text{-}A}$ reconstructed by the NODDI fitting and DIFFnet$_{NODDI}$ using the five different numbers of the diffusion gradients. (b) NRMSEs in the DIFFnet$_{NODDI}$ and NODDI results. The reference NRMSEs was the NODDI fitting results using the 104 diffusion gradient directions.

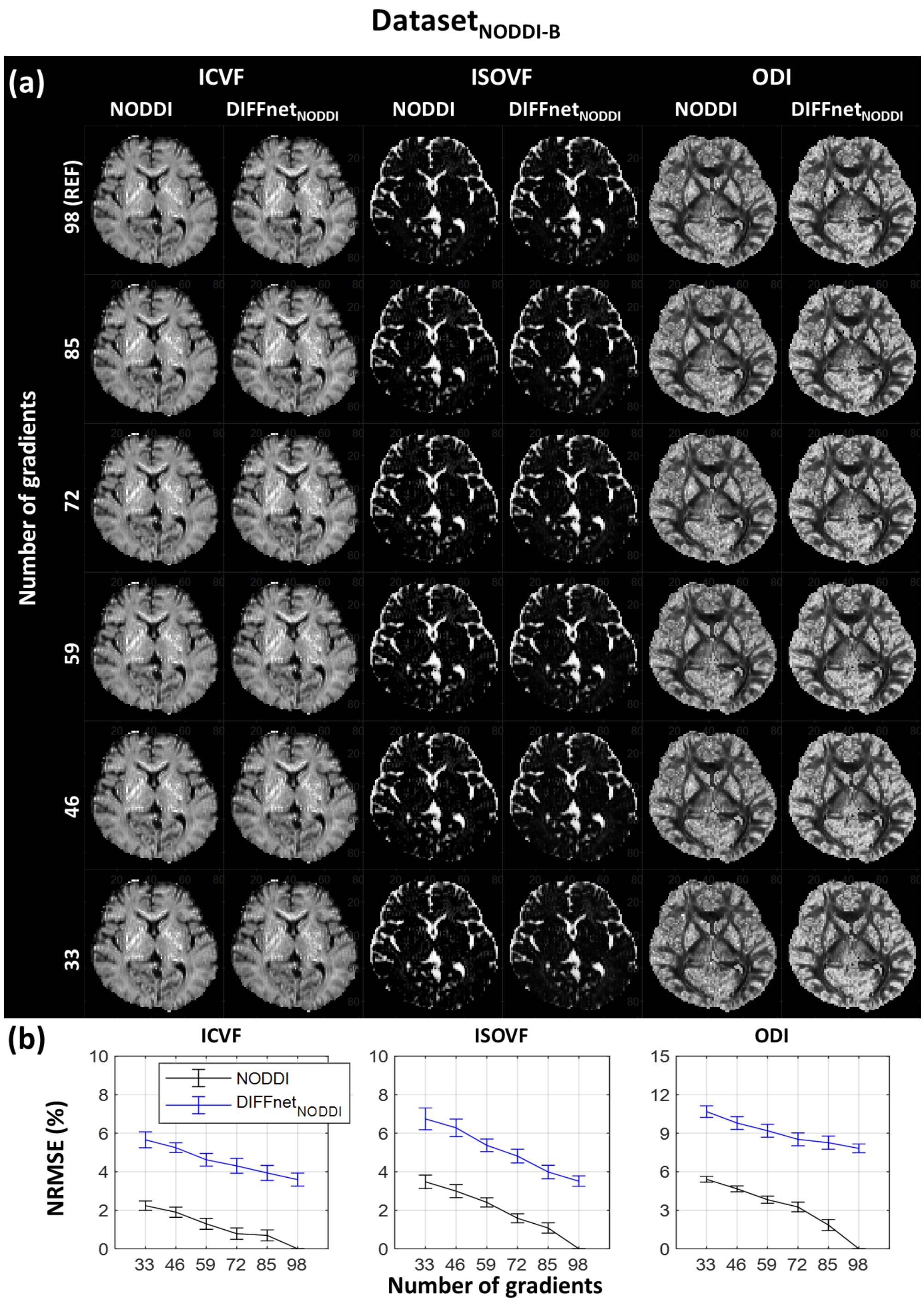

**Fig S5.** (a) NODDI parameter maps of Dataset$_{NODDI-B}$ reconstructed by the NODDI fitting and DIFFnet$_{NODDI}$ using the five different numbers of the diffusion gradients. (b) NRMSEs in the DIFFnet$_{NODDI}$ and NODDI results. The reference maps was the NODDI fitting results using the 98 gradient directions.

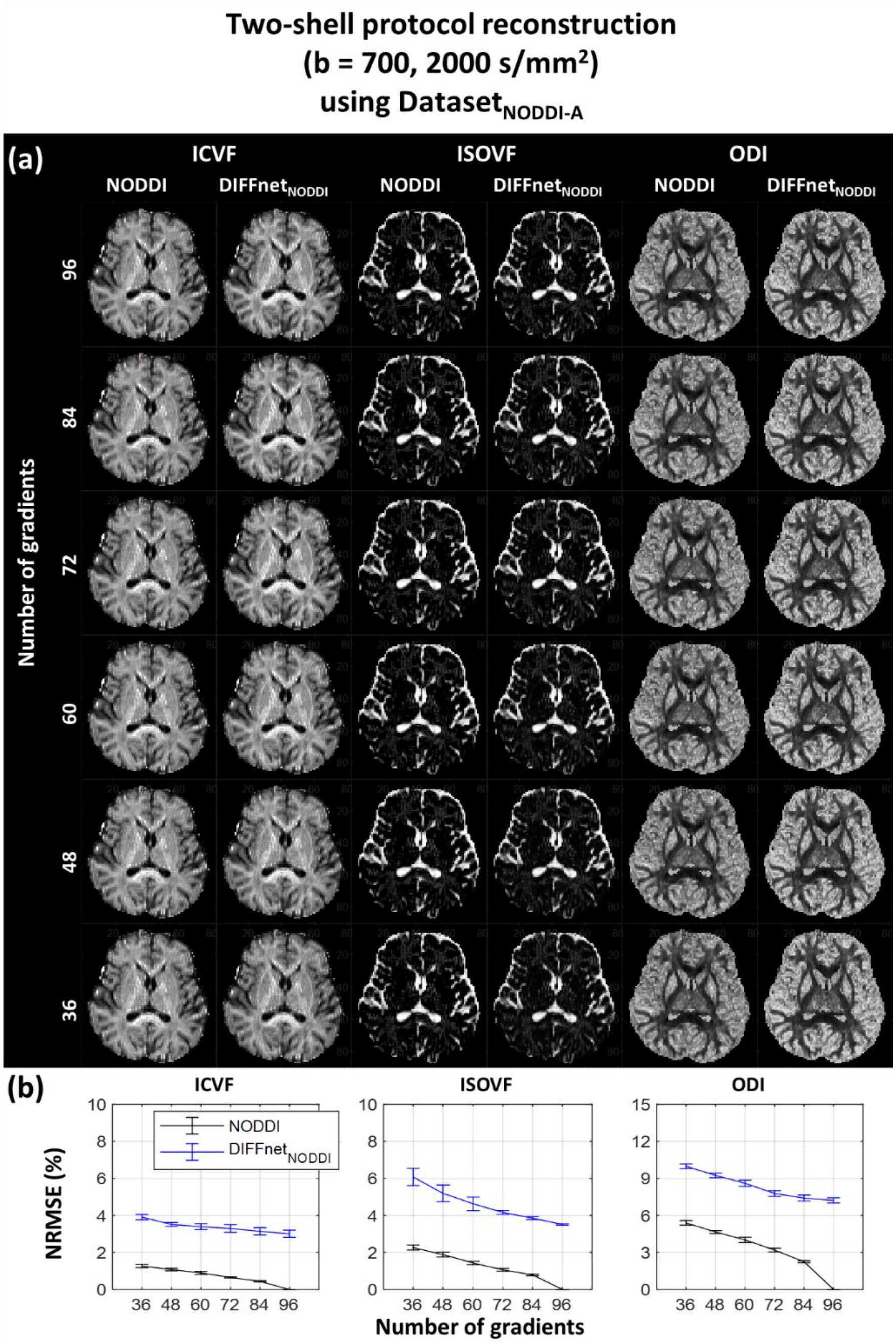

**Fig S6.** (a) NODDI parameter maps of Dataset$_{NODDI\text{-}A}$ reconstructed by the NODDI fitting and DIFFnet$_{NODDI}$ using the two-shell protocol (b = 700 and 2000 s/mm$^2$), with the six different numbers of the diffusion gradients. (b) NRMSEs in the DIFFnet$_{NODDI}$ and NODDI results. The reference for NRMSE was the NODDI results using the 96 gradient directions (b = 700 s/mm$^2$ with 32 directions; b = 2000 s/mm$^2$ with 64 directions).

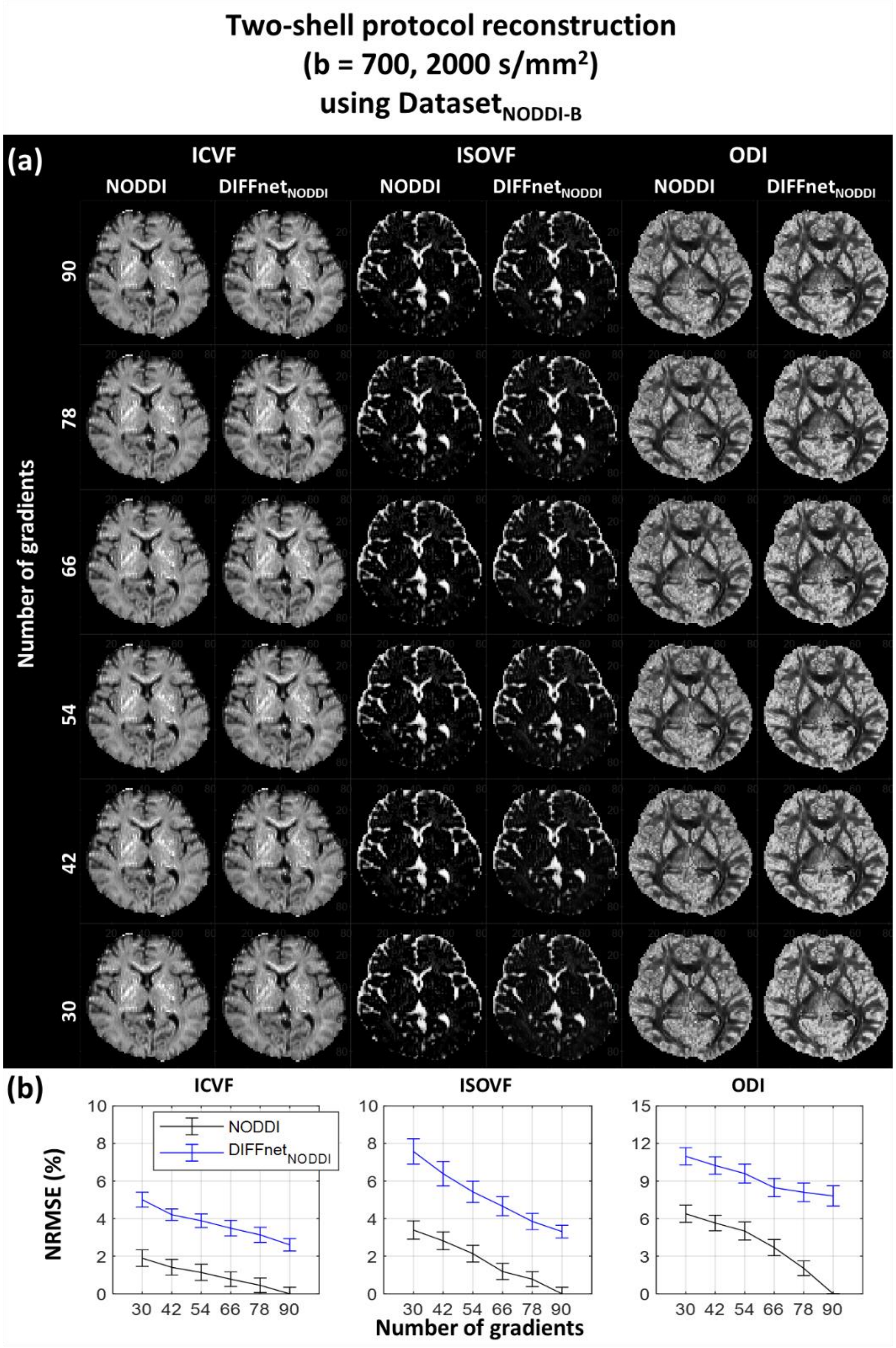

**Fig S7.** (a) NODDI parameter maps of Dataset$_{NODDI-B}$ reconstructed by the NODDI fitting and DIFFnet$_{NODDI}$ using the two-shell protocol (b = 700 and 2000 s/mm$^2$), with the six different numbers of the diffusion gradients. (b) NRMSEs in the DIFFnet$_{NODDI}$ and NODDI results. The reference for NRMSE was the NODDI results using the 90 gradient directions (b = 700 s/mm$^2$ with 30 directions; b = 2000 s/mm$^2$ with 60 directions).

TABLE SI

GRADIENT DIRECTIONS OF THE TEST DATASETS

| Dataset$_{DTI-A}$ & Dataset$_{NODDI-A}$ | | | | Dataset$_{DTI-B}$ | | | | Dataset$_{NODDI-B}$ | | | |
|---|---|---|---|---|---|---|---|---|---|---|---|
| b-value (s/mm$^2$) | x | y | z | b-value (s/mm$^2$) | x | y | z | b-value (s/mm$^2$) | x | y | z |
| 300 | 0.392 | -0.033 | -0.919 | 1000 | 0.802 | -0.064 | -0.593 | 300 | -0.594 | -0.804 | 0.025 |
| | -0.391 | -0.421 | -0.819 | | -0.468 | -0.55 | -0.692 | | -0.014 | -0.734 | 0.68 |
| | -0.592 | -0.804 | 0.067 | | 0.601 | -0.493 | -0.629 | | 0.761 | -0.646 | -0.062 |
| | 0.028 | -0.73 | 0.683 | | -0.801 | 0.223 | -0.556 | | 0.967 | 0.144 | -0.21 |
| | 0.756 | -0.646 | -0.105 | | 0.38 | 0.404 | -0.832 | | 0.448 | -0.033 | -0.893 |
| | 0.954 | 0.143 | -0.263 | | -0.103 | 0.449 | -0.888 | | 0.125 | -0.919 | -0.373 |
| | 0.101 | -0.922 | -0.374 | | 0.749 | 0.363 | -0.555 | | -0.337 | -0.419 | -0.843 |
| | -0.754 | 0.244 | -0.61 | | 0.479 | -0.041 | -0.877 | | -0.709 | 0.244 | -0.661 |
| 700 | 0.803 | -0.064 | -0.593 | | 0.232 | -0.429 | -0.873 | 700 | 0.542 | 0.726 | -0.423 |
| | -0.468 | -0.551 | -0.691 | | -0.207 | -0.304 | -0.93 | | 0.133 | 0.965 | -0.224 |
| | 0.601 | -0.494 | -0.628 | | 0.119 | 0.751 | -0.649 | | -0.423 | -0.549 | -0.721 |
| | -0.802 | 0.223 | -0.554 | | -0.437 | 0.091 | -0.895 | | 0.673 | 0.197 | 0.713 |
| | 0.381 | 0.405 | -0.832 | | -0.499 | 0.515 | -0.697 | | 0.638 | -0.493 | -0.592 |
| | -0.103 | 0.449 | -0.887 | | 0.058 | 0.049 | -0.997 | | -0.763 | 0.223 | -0.607 |
| | 0.749 | 0.364 | -0.554 | | 0.519 | 0.854 | -0.029 | | -0.013 | 0.734 | 0.679 |
| | 0.48 | -0.041 | -0.877 | | 0.515 | 0.727 | -0.453 | | 0.431 | 0.403 | -0.807 |
| | 0.232 | -0.429 | -0.873 | | 0.118 | 0.966 | -0.229 | | -0.932 | -0.161 | -0.324 |
| | -0.207 | -0.305 | -0.93 | | -0.952 | -0.161 | -0.261 | | -0.047 | 0.448 | -0.893 |
| | 0.119 | 0.752 | -0.649 | | -0.761 | -0.567 | -0.314 | | -0.739 | -0.567 | -0.364 |
| | -0.437 | 0.092 | -0.895 | | -0.698 | 0.654 | -0.292 | | -0.677 | 0.654 | -0.338 |
| | -0.499 | 0.515 | -0.697 | | -0.945 | 0.294 | -0.142 | | -0.934 | 0.293 | -0.204 |
| | 0.058 | 0.049 | -0.997 | | -0.354 | 0.934 | -0.049 | | -0.349 | 0.934 | -0.074 |
| | 0.519 | 0.854 | -0.027 | | -0.681 | 0.715 | 0.161 | | -0.689 | 0.715 | 0.115 |
| | 0.516 | 0.727 | -0.453 | | 0.891 | -0.361 | -0.277 | | 0.781 | 0.363 | -0.508 |
| | 0.118 | 0.966 | -0.228 | | -0.379 | -0.846 | -0.375 | | 0.906 | -0.36 | -0.223 |
| | -0.952 | -0.161 | -0.26 | | -0.282 | 0.833 | -0.476 | | -0.354 | -0.845 | -0.4 |
| | -0.761 | -0.567 | -0.314 | | 0.842 | 0.525 | -0.122 | | -0.251 | 0.832 | -0.495 |
| | -0.698 | 0.654 | -0.291 | | 0.979 | 0.099 | -0.18 | | 0.848 | 0.525 | -0.072 |
| | -0.946 | 0.293 | -0.141 | | 0.004 | 0.977 | 0.214 | | 0.532 | -0.041 | -0.846 |
| | -0.353 | 0.934 | -0.048 | | -0.348 | 0.817 | 0.46 | | 0.286 | -0.428 | -0.858 |
| | -0.68 | 0.715 | 0.162 | | 0.715 | 0.197 | 0.671 | | -0.148 | -0.304 | -0.941 |
| | 0.891 | -0.36 | -0.276 | | 0.03 | 0.732 | 0.681 | | 0.988 | 0.099 | -0.122 |
| | -0.379 | -0.846 | -0.374 | | | | | | 0.159 | 0.75 | -0.642 |
| | -0.282 | 0.834 | -0.474 | | | | | | -0.009 | 0.978 | 0.21 |
| | 0.842 | 0.525 | -0.121 | | | | | | -0.376 | 0.818 | 0.435 |
| | 0.979 | 0.099 | -0.179 | | | | | | -0.378 | 0.091 | -0.921 |
| | 0.004 | 0.977 | 0.215 | | | | | | -0.454 | 0.513 | -0.729 |
| | -0.348 | 0.816 | 0.461 | | | | | | 0.12 | 0.05 | -0.992 |
| | 0.714 | 0.197 | 0.671 | | | | | 2000 | -1 | 0.027 | 0 |
| | 0.029 | 0.731 | 0.682 | | | | | | 0.047 | 0.014 | -0.999 |
| 2000 | -0.016 | 0.015 | -1 | | | | | | 0.813 | 0.38 | 0.441 |
| | 0.308 | -0.091 | -0.947 | | | | | | 0.233 | -0.891 | -0.39 |
| | -0.329 | 0.035 | -0.944 | | | | | | 0.589 | -0.252 | -0.768 |
| | -0.056 | 0.318 | -0.947 | | | | | | -0.112 | -0.936 | -0.334 |
| | 0.065 | -0.296 | -0.953 | | | | | | 0.818 | -0.549 | 0.172 |
| | -0.24 | -0.286 | -0.928 | | | | | | 0.366 | -0.091 | -0.926 |
| | 0.277 | 0.218 | -0.936 | | | | | | 0.27 | 0.505 | -0.82 |
| | 0.798 | 0.137 | -0.587 | | | | | | 0.735 | -0.666 | -0.124 |
| | 0.27 | -0.713 | -0.647 | | | | | | -0.284 | -0.557 | -0.78 |
| | -0.646 | -0.682 | -0.343 | | | | | | -0.268 | 0.034 | -0.963 |
| | 0.208 | -0.892 | -0.402 | | | | | | 0.317 | 0.864 | -0.392 |
| | 0.54 | -0.253 | -0.803 | | | | | | -0.55 | 0.742 | 0.383 |
| | -0.133 | -0.937 | -0.324 | | | | | | -0.072 | 0.59 | -0.804 |
| | 0.218 | 0.506 | -0.835 | | | | | | 0.929 | 0.086 | 0.359 |
| | -0.333 | -0.558 | -0.76 | | | | | | 0.004 | 0.317 | -0.949 |
| | 0.292 | 0.865 | -0.409 | | | | | | -0.107 | 0.825 | -0.556 |
| | -0.122 | 0.591 | -0.798 | | | | | | 0.592 | -0.79 | 0.158 |
| | -0.142 | 0.826 | -0.546 | | | | | | 0.765 | 0.444 | -0.467 |
| | 0.735 | 0.445 | -0.512 | | | | | | 0.661 | 0.75 | 0.035 |
| | -0.613 | -0.503 | -0.61 | | | | | | -0.572 | -0.502 | -0.648 |
| | -0.832 | 0.382 | -0.403 | | | | | | -0.804 | 0.381 | -0.457 |
| | -0.371 | 0.345 | -0.862 | | | | | | -0.316 | 0.344 | -0.884 |
| | -0.763 | -0.186 | -0.618 | | | | | | -0.722 | -0.186 | -0.666 |
| | 0.927 | -0.117 | -0.356 | | | | | | 0.125 | -0.295 | -0.947 |
| | -0.063 | -0.797 | -0.601 | | | | | | 0.947 | -0.117 | -0.3 |
| | 0.769 | -0.194 | -0.608 | | | | | | -0.026 | -0.795 | -0.606 |
| | -0.523 | -0.249 | -0.815 | | | | | | 0.805 | -0.194 | -0.561 |
| | -0.379 | -0.767 | -0.517 | | | | | | 0.454 | -0.883 | -0.115 |
| | -0.638 | 0.377 | -0.671 | | | | | | -0.846 | -0.51 | 0.154 |
| | 0.306 | -0.465 | -0.831 | | | | | | -0.41 | -0.883 | -0.229 |
| | 0.534 | 0.371 | -0.76 | | | | | | -0.47 | -0.249 | -0.847 |
| | -0.408 | 0.823 | -0.395 | | | | | | -0.345 | -0.766 | -0.542 |
| | -0.837 | 0.105 | -0.538 | | | | | | -0.594 | 0.376 | -0.711 |
| | 0.928 | 0.216 | -0.303 | | | | | | -0.946 | 0.217 | -0.239 |
| | 0.458 | 0.643 | -0.614 | | | | | | 0.356 | -0.464 | -0.811 |
| | 0.612 | 0.713 | -0.342 | | | | | | -0.18 | -0.285 | -0.941 |
| | 0.582 | 0.069 | -0.811 | | | | | | 0.579 | 0.37 | -0.726 |
| | -0.652 | 0.631 | -0.419 | | | | | | -0.381 | 0.823 | -0.421 |
| | -0.414 | 0.62 | -0.666 | | | | | | -0.8 | 0.105 | -0.591 |
| | 0.577 | -0.523 | -0.627 | | | | | | 0.945 | 0.216 | -0.247 |
| | 0.793 | -0.456 | -0.403 | | | | | | 0.929 | -0.363 | -0.069 |
| | -0.609 | 0.077 | -0.79 | | | | | | -0.133 | 0.985 | 0.107 |
| | -0.02 | -0.577 | -0.816 | | | | | | -0.427 | -0.898 | 0.109 |
| | 0.138 | 0.748 | -0.65 | | | | | | 0.495 | 0.642 | -0.586 |
| | -0.279 | 0.949 | -0.148 | | | | | | 0.632 | 0.712 | -0.305 |
| | -0.997 | 0.027 | 0.066 | | | | | | 0.63 | 0.069 | -0.774 |
| | 0.827 | -0.549 | 0.123 | | | | | | -0.155 | -0.988 | 0.003 |
| | 0.726 | -0.666 | -0.168 | | | | | | -0.624 | 0.631 | -0.461 |
| | 0.601 | -0.79 | 0.124 | | | | | | 0.334 | 0.218 | -0.917 |
| | 0.661 | 0.75 | -0.004 | | | | | | -0.828 | -0.533 | -0.175 |
| | 0.446 | -0.884 | -0.141 | | | | | | -0.371 | 0.619 | -0.692 |
| | -0.835 | -0.51 | 0.209 | | | | | | 0.614 | -0.522 | -0.591 |
| | -0.424 | -0.883 | -0.201 | | | | | | 0.817 | -0.455 | -0.355 |
| | -0.96 | 0.217 | -0.177 | | | | | | -0.557 | 0.077 | -0.827 |
| | 0.923 | -0.363 | -0.124 | | | | | | 0.959 | 0.267 | 0.089 |
| | -0.126 | 0.985 | 0.117 | | | | | | 0.025 | 0.955 | -0.295 |
| | -0.419 | -0.897 | 0.138 | | | | | | 0.031 | -0.576 | -0.817 |
| | -0.155 | -0.988 | 0.016 | | | | | | 0.179 | 0.746 | -0.641 |
| | -0.838 | -0.533 | -0.12 | | | | | | | | |
| | 0.963 | 0.267 | 0.032 | | | | | | | | |
| | 0.007 | 0.956 | -0.294 | | | | | | | | |
| | 0.839 | 0.379 | 0.391 | | | | | | | | |
| | -0.524 | 0.742 | 0.419 | | | | | | | | |
| | 0.949 | 0.086 | 0.302 | | | | | | | | |

## Supplementary information reference